\documentclass[12pt]{article}
 \usepackage[dvips,final]{graphicx}
  \usepackage{amsmath,amssymb,bm,pifont}

\graphicspath{{./Figs-AB/}}

\newcommand{\Ds}{\displaystyle}                                    
\newcommand{\gev}[1]{\relax\ifmmode{\text{GeV}^{#1}}               
                     \else{{GeV}$^{#1}${ }}\fi}                    
\newcommand{\Gev}{\relax\ifmmode{\text{GeV}}\else{{ GeV}{ }}\fi}   
 \newcommand{\Mev}{\relax\ifmmode{\text{MeV}}\else{{MeV}{ }}\fi}   
\newcommand{\bmp}[1]{\begin{minipage}{#1}}\newcommand{\emp}{\end{minipage}}%
\def\muF{\relax\ifmmode\mu_\text{F}^2\else{$\mu_\text{F}^2${ }}\fi}
\def\muR{\relax\ifmmode\mu_\text{R}^2\else{$\mu_\text{R}^2${ }}\fi}
\def\muO{\relax\ifmmode{\mu_{0}^{2}}\else{$\mu_{0}^{2}${ }}\fi}    
\def\MS{$\overline{\text{MS}\vphantom{^1}}${ }}                    
\def\as{\relax\ifmmode \alpha_s\else{$ \alpha_s${ }}\fi}           
\def\asb{\relax\ifmmode \bar{\alpha}_s\else{$ \bar{\alpha}_s${ }}\fi}
 \def\abar{\relax\ifmmode{\bar{a}}\else{$\bar{a}${ }}\fi}          
  \def\acal{\relax\ifmmode{\cal A}\else{${\cal A}${ }}\fi}         
\newcommand\convo[1]{\mathop{\otimes}\limits_{#1}}                 

\begin{document}

 \begin{center}
 \textbf{\LARGE  Fractional APT in QCD\\[1mm]
  in the Euclidean and Minkowski regions}\\[5mm]
 
 Alexander~P.~Bakulev\footnote{E-mail: bakulev@theor.jinr.ru}\\[3mm]

\textit{Bogoliubov  Lab. of Theoretical Physics,\\
        Joint Institute for Nuclear Research,
        141980, Dubna, Russia}\\[0.9cm]
\end{center}\vspace*{-5mm}

\begin{abstract}
 We describe the development of Analytic Perturbation Theory (APT) in QCD,
called Fractional APT (FAPT), 
which has been suggested to apply the renormalization group evolution 
and QCD factorization technique in the framework of APT.
\end{abstract}

\section{Basics of APT in QCD}
In the standard QCD Perturbation Theory (PT) we have:
 \begin{itemize}
   \item[\ding{52}] the Renormalization Group (RG) equation 
   $da_s(L)/dL = -a_s^2-c_1\,a_s^3-\ldots$
   for the effective coupling $\alpha_s(\mu^2)=(4\pi/b_0)\,a_s(L)$ 
   with $L=\ln(\mu^2/\Lambda^2)$;   
   \item[\ding{52}] the one-loop solution generates Landau pole singularity: 
     $a_s(L) = 1/L$;
   \item[\ding{52}] the two-loop solution generates square-root singularity: 
     $a_s(L) \sim 1/\sqrt{L+c_1\ln c_1}$;
   \item[\ding{52}] PT series is a series in powers of effective coupling:
     $D(L) = 1 + d_1 a_s(L)+ d_2 a_s^2(L)+ \ldots$
  \end{itemize}
In the Analytic Perturbation Theory (APT) we have:
  \begin{itemize}
   \item[\ding{52}] different effective couplings in 
         Minkowskian (Radyushkin \cite{Rad82}, and Krasnikov and Pivovarov \cite{KP82})
         and Euclidean (Shirkov and Solovtsov \cite{SS96}) regions;
   \item[\ding{52}] APT is based on the RG and causality 
   that guaranties standard perturbative UV asymptotics 
   and spectral properties;
   \item[\ding{52}] in Euclidean domain, 
    $\Ds-q^2=Q^2$, $\Ds L=\ln Q^2/\Lambda^2$,
    APT generates the following set of images for the effective coupling
    and its $n$-th powers,  
    $\Ds\left\{{\cal A}_n(L)\right\}_{n\in\mathbb{N}}$;
   \item[\ding{52}] in Minkowskian domain,
    $\Ds q^2=s$, $\Ds L_s=\ln s/\Lambda^2$,
    APT generates another set of images for the effective coupling
    and its $n$-th powers,
    $\Ds\left\{{\mathfrak A}_n(L_s)\right\}_{n\in\mathbb{N}}$;
   \item[\ding{52}] PT power series 
   $\Ds\sum\limits_{m}d_m a_s^m(Q^2)$ 
   transforms into non-power series 
   $\Ds\sum\limits_{m}d_m {\cal A}_{m}(Q^2)$ in APT,
   where $d_m$ are \textsl{numbers} in \MS-scheme.
   \end{itemize}
By the analytization in APT for an observable $f(Q^2)$
we mean the ``K\"allen--Lehman'' representation
 \begin{eqnarray}
  \left[f(Q^2)\right]_\text{an}
   = \int_0^{\infty}\!
      \frac{\rho_f(\sigma)}
         {\sigma+Q^2-i\epsilon}\,
       d\sigma
 \end{eqnarray}
with the spectral density 
$\rho_f(\sigma)=\textbf{Im}\,\big[f(-\sigma)\big]/\pi$.
Then in the one-loop approximation (note pole remover $(e^L-1)^{-1}$ in (\ref{eq:A_1}))
\begin{eqnarray}
 \label{eq:A_1}
 {\cal A}_1(Q^2)
  &=& \int_0^{\infty}\!\frac{\rho(\sigma)}{\sigma+Q^2}\,d\sigma\
   =\ \frac{1}{L} - \frac{1}{e^L-1}\,,~\\
 \label{eq:U_1}  
 {\mathfrak A}_1(s) 
  &=& \int_s^{\infty}\!\frac{\rho(\sigma)}{\sigma}\,d\sigma\
   =\ \frac{1}{\pi}\,\arccos\frac{L_s}{\sqrt{\pi^2+L_s^2}}\,,~
\end{eqnarray}
whereas analytic images of the higher powers ($n\geq2, n\in\mathbb{N}$) are:
\begin{eqnarray}\!\!\!\!
 {\cal A}_n(Q^2)
 \!\!&\!=\!\!&\!\!
  \int_0^{\infty}\!\frac{\rho_n(\sigma)}{\sigma+Q^2}\,d\sigma
  \!=\! \frac{1}{(n-1)!}\left( -\frac{d}{d L}\right)^{n-1}\!\!\!\!\!
     {\cal A}_{1}(L)\,,~
 \\
 {\mathfrak A}_n(s)
 \!\!&\!=\!\!&\!\!
  \int_s^{\infty}\!\frac{\rho_n(\sigma)}{\sigma}\,d\sigma
  \!=\! \frac{1}{(n-1)!}\left(-\frac{d}{d L}\right)^{n-1}\!\!\!\!\!
    {\mathfrak A}_{1}(L)\,.
\end{eqnarray}

\section{Problems of APT and their resolution in FAPT}
In the standard QCD PT we have not only power series
 $F(L)= \sum\limits_{m}f_m\,a_s^m(L)$,
but also:
 \begin{itemize} 
  \item[\ding{52}] the factorization procedure in QCD  
     gives rise to the appearance of logarithmic factors of the type: 
     $a_s^\nu(L)\,L$;~\footnote{%
     First indication that a special ``analytization'' procedure
is needed to handle these logarithmic terms appeared in~\cite{KS01},
where it has been suggested that one should demand 
the analyticity of the partonic amplitude as a \textit{whole}.}
  \item[\ding{52}] the RG evolution generates evolution factors of the type: 
     $B(Q^2)=\left[Z(Q^2)/Z(\mu^2)\right]\,B(\mu^2)$, 
     which reduce in the one-loop approximation to
     $Z \sim a_s^\nu(L)$ with $\nu=\gamma_0/(2b_0)$ 
     being a fractional number;
  \item[\ding{52}] the RG in the two-loop approximation for the coupling  
   $\Ds\to\left[a_s(L)\right]^\nu\ln\left(a_s(L)\right)$.
 \end{itemize}
That means we need to think how to obtain analytization of new functions:
$\Ds\left(a_s\right)^\nu,\ \left(a_s\right)^\nu\ln(a_s)$,
$\Ds\left(a_s\right)^\nu L^m$, $\ldots$\,.

 Let us first do it for the one-loop APT. 
Here we have a very nice recursive relation (\ref{eq:A_1}).
We will use it to construct analytic images of fractional powers 
of QCD effective coupling in the Euclidean (FAPT) and Minkowskian (MFAPT) domains.
Consider the Laplace transform
\begin{eqnarray}
 \label{eq:A1_Lap}
  {{\cal A}_{1}(L)\choose {\mathfrak A}_{1}(L)}
    = \int_0^{\infty}\!
      {\tilde{{\cal A}}_{1}(t)\choose {\mathfrak A}_{1}(t)}\,
      e^{-L t} dt\,,
\end{eqnarray}
which is well defined for all $L>0$.
Then 
\begin{eqnarray}
 {{\cal A}_{n}(L)\choose {\mathfrak A}_{n}(L)}
    = \int_0^{\infty}\!
      {\tilde{{\cal A}}_{1}(t)\choose {\mathfrak A}_{1}(t)}
       \left[\frac{t^{n-1}}{(n-1)!}\right]\,
        e^{-L t} dt\,.
\end{eqnarray}
Moreover, we can define for all $\nu\in\mathbb{R}$
\begin{eqnarray}\label{eq:A.nu.Laplace}
{{\cal A}_{\nu}(L)\choose {\mathfrak A}_{\nu}(L)}
    = \int_0^{\infty}\!
      {\tilde{{\cal A}}_{1}(t)\choose {\mathfrak A}_{1}(t)}
       \left[\frac{t^{\nu-1}}{(\nu-1)!}\right]\,
        e^{-L t} dt\,.
\end{eqnarray}
The only things one needs to know are $\tilde{{\cal A}}_{1}(t)$
and $\tilde{{\mathfrak A}}_{1}(t)$.
Eqs.\ (\ref{eq:A_1}) and (\ref{eq:U_1}) produce the answer:
$\Ds\tilde{{\cal A}}_{1}(t)= 1 - \sum_{m=1}^{\infty}\delta(t-m)$
and $\Ds\tilde{{\mathfrak A}}_{1}(t)= \left[\frac{\textbf{sin}\pi t}{\pi t}\right]$.
This allows us to obtain explicit expressions for
${\cal A}_{\nu}(L)$ ($L=\ln Q^2/\Lambda^2$)
and ${\mathfrak A}_{\nu}(L)$ ($L=\ln s/\Lambda^2$)
using Eq.\ (\ref{eq:A.nu.Laplace}):
\begin{eqnarray}
 {\cal A}_{\nu}(L) 
  = \frac{1}{L^\nu} 
  - \frac{F(e^{-L},1-\nu)}{\Gamma(\nu)}\,;\quad
 {\mathfrak A}_{\nu}(L) 
  = \frac{\text{sin}\left[(\nu -1)\arccos\left(L/\sqrt{\pi^2+L^2}\right)\right]}
         {\pi(\nu -1) \left(\pi^2+L^2\right)^{(\nu-1)/2}}\,.~
\end{eqnarray}
Here $F(z,\nu)$ is reduced Lerch transcendental function.
It is an analytic function in $\nu$.
Interesting to note that ${\cal A}_\nu(L)$ appears to be 
an entire function in $\nu$, 
whereas ${\mathfrak A}_{\nu}(L)$ 
is determined completely in terms of elementary functions.
These expressions can be analytically continued to negative values
of $L$, 
though in derivation we assume $L>0$.

Let us discuss the main properties of ${\cal A}_{\nu}(L)$ ($L=\ln Q^2/\Lambda^2$)
and ${\mathfrak A}_{\nu}(L)$ ($L=\ln s/\Lambda^2$):
\begin{figure}[t]
 \centerline{\includegraphics[width=0.45\textwidth]{
    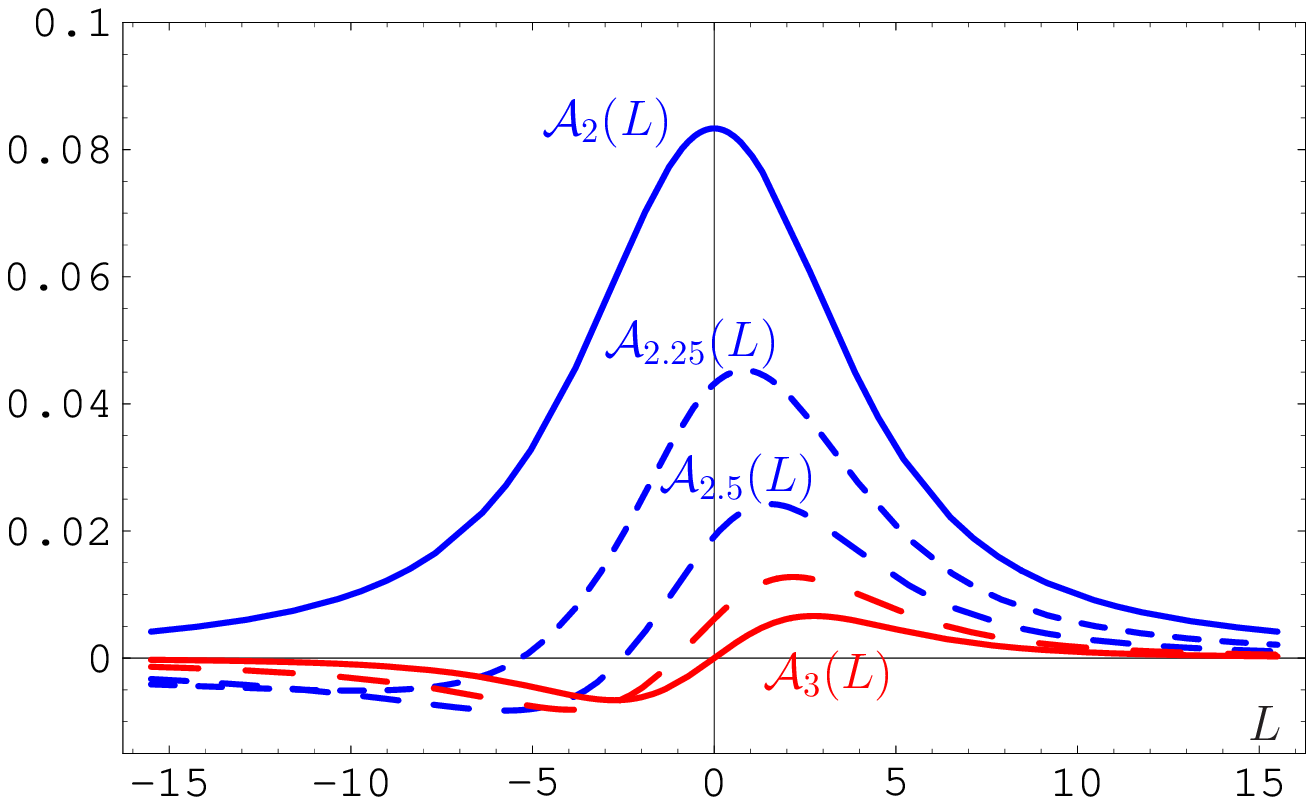}~~~%
             \includegraphics[width=0.45\textwidth]{
    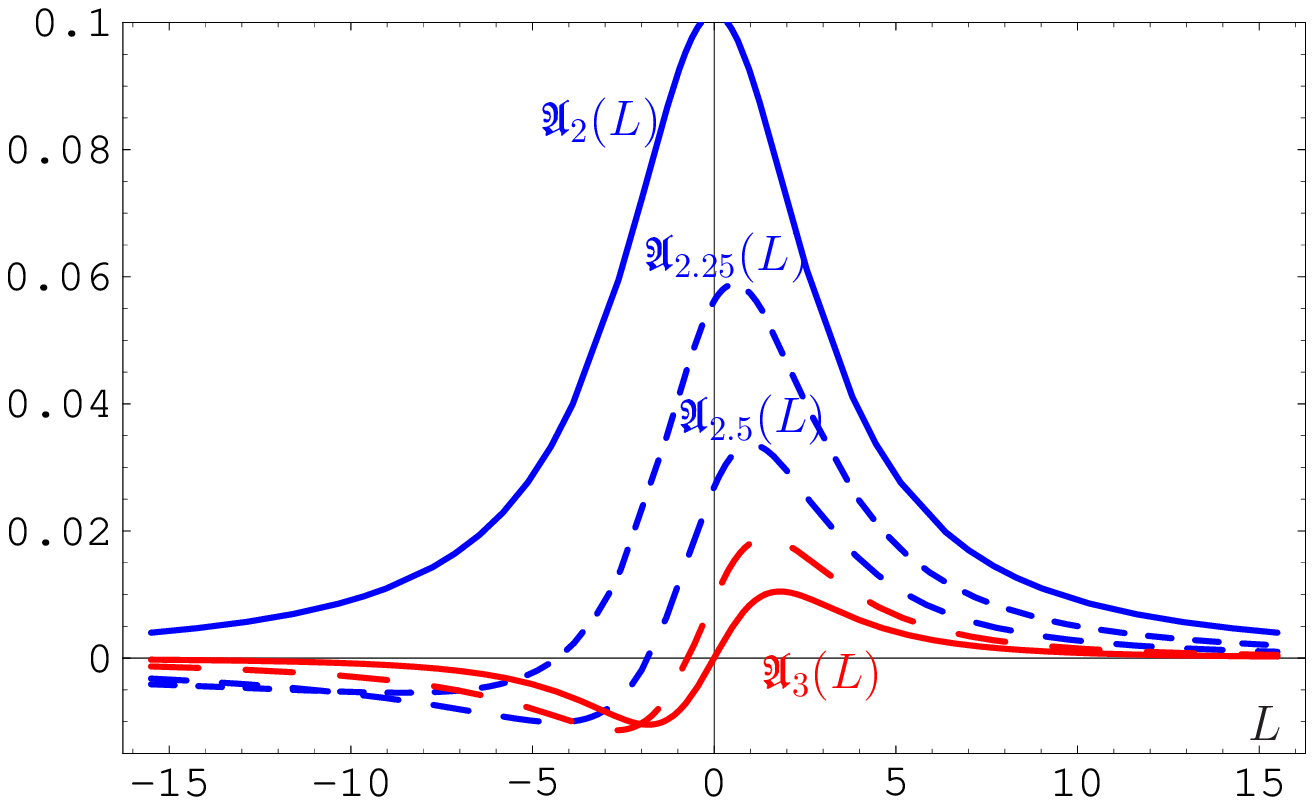}}
  \caption{Graphics of ${\cal A}_{\nu}(L)$ (left panel)
  and ${\mathfrak A}_{\nu}(L)$ (right panel)
  for fractional $\nu\in\left[2,3\right]$.
  \label{fig:U23_A23}}
\end{figure} 
These couplings have the following properties:\\
\ding{202}\ $\Ds{\cal A}_{0}(L)={\mathfrak A}_{0}(L)=1\vphantom{^{\int}_{|}}$;\\
\ding{203}\ $\Ds{\cal A}_{-m}(L)=L^m$ for $m\in\mathbb{N}$;~
    $\Ds{\mathfrak A}_{-1}(L)=L$,~
    $\Ds{\mathfrak A}_{-2}(L)=L^2-\frac{\pi^2}{3}$,~
    $\Ds{\mathfrak A}_{-3}(L)=L^3-\pi^2L\,,\,\ldots\vphantom{^{\int}_{|}}$;\\
\ding{204}\ $\Ds{{\cal A}_{m}(L)\vphantom{^{\int}}\choose{\mathfrak A}_{m}(L)\vphantom{_{|}}}
         \!\!=\!\!(-1)^{m}{{\cal A}_{m}(-L)\choose{\mathfrak A}_{m}(-L)}$ 
             for $m\geq2\,, m\in\mathbb{N}$;\\
\ding{205}\ $\Ds{\cal A}_{m}(\pm\infty)={\mathfrak A}_{m}(\pm\infty)=0$
                     for $m\geq2\,,\ m\in\mathbb{N}\vphantom{^{\int}_{\big|}}$;\\
\ding{206}\ $\Ds{\cal D}^{k}{{\cal A}_{\nu}\vphantom{^{\int}}\choose{\mathfrak A}_{\nu}\vphantom{_{|}}} 
             \equiv \frac{d^k}{d \nu^k}{{\cal A}_{\nu}\choose{\mathfrak A}_{\nu}}
             = \left[\frac{d^k}{d \nu^k}a^\nu\right]_\text{an}
             = \left[a^\nu\ln^k(a)\right]_\text{an}$;\\
\ding{207}\ $\Ds{\cal A }_{\nu}(L)
             = \frac{-1}{\Gamma(\nu)}
                \sum_{r=0}^{\infty}\zeta(1-\nu-r)\frac{(-L)^{r}}{r!}\vphantom{^{\int}_{|}}$ 
             for $|L|\!<\!2\pi$. The convergence of this expansion is very fast.               
\begin{figure}[bh]
 \centerline{\includegraphics[width=0.45\textwidth]{
    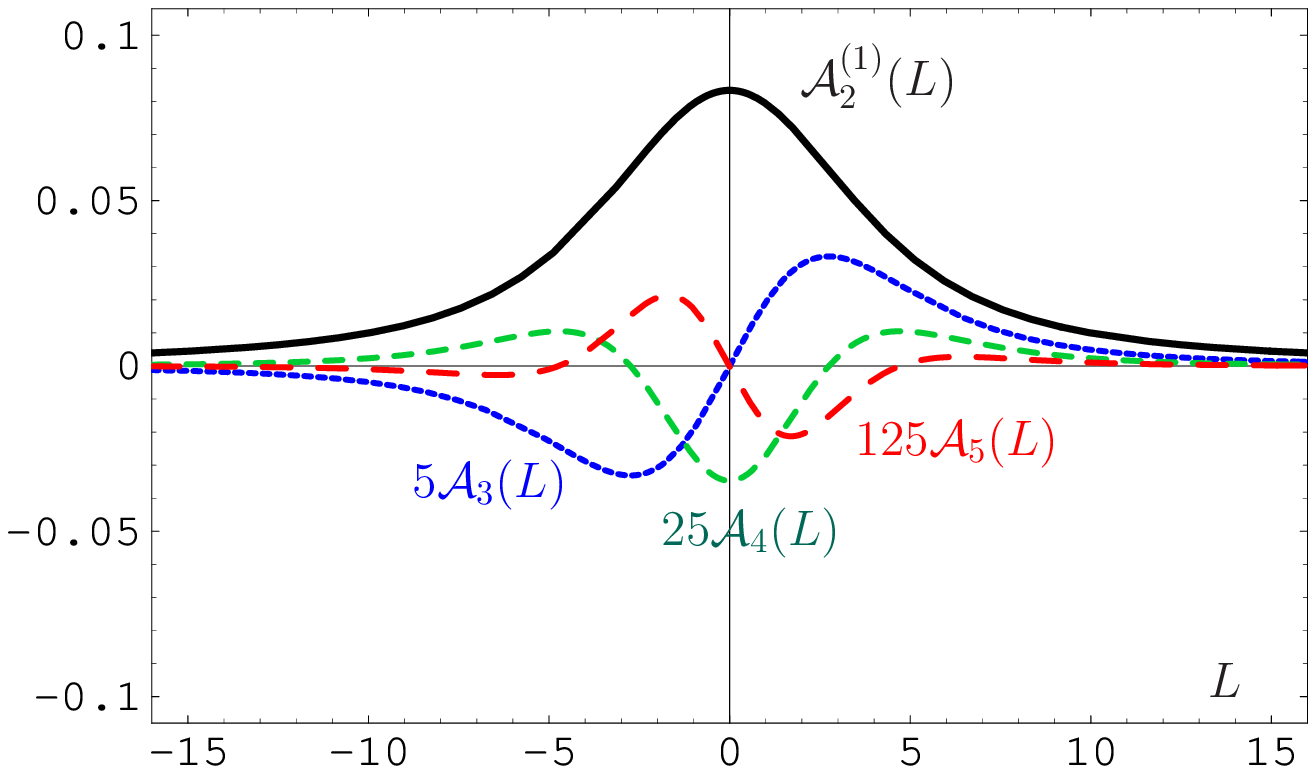}~~~%
             \includegraphics[width=0.45\textwidth]{
    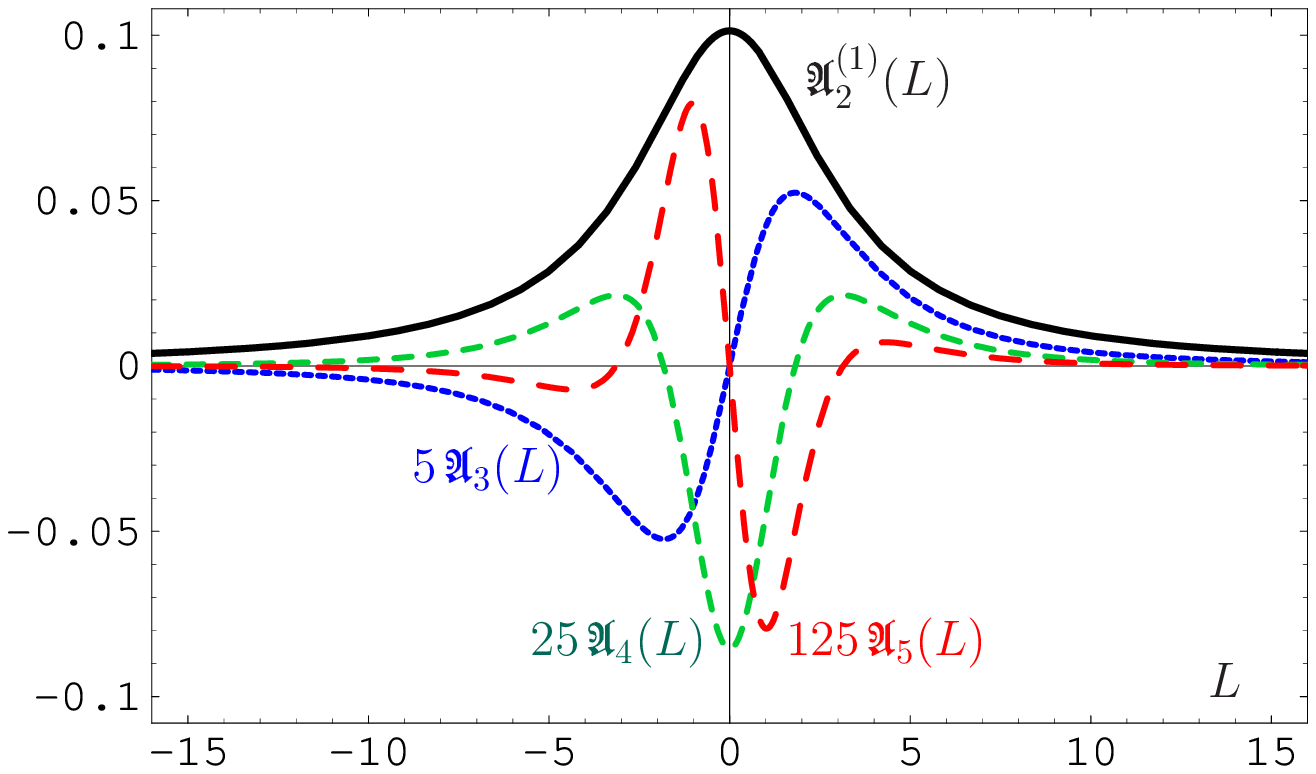}}
  \caption{Graphics of ${\cal A}_{\nu}(L)$ (left panel)
  and ${\mathfrak A}_{\nu}(L)$ (right panel)
  for integer $\nu=2, 3, 4, 5$. In order to show all curves 
  on the same panel we scale different curves by factors $5^{\nu-2}$. 
  \label{fig:UA_2345}}
\end{figure}
We display graphics of ${\cal A}_{\nu}(L)$ and ${\mathfrak A}_{\nu}(L)$
in Fig.\ \ref{fig:U23_A23}:
one can see here a kind of distorting mirror on both panels.
Next, in Fig.\ \ref{fig:UA_2345} we show graphics for $\nu=2, 3, 4, 5$.
Here we can trace the partial values
\begin{eqnarray*}
 {\cal A }_{2}(0) = \frac{1}{12}\,,~~
 {\cal A }_{4}(0) = \frac{-1}{720}\,,~~
 {\cal A }_{3}(0) = {\cal A }_{5}(0)=0\,;~~\\
 {\mathfrak A }_{2}(0) = \frac{1}{\pi^2}\,,~~
 {\mathfrak A }_{4}(0) =-\frac{1}{3\pi^4}\,,~~
 {\mathfrak A }_{3}(0) = {\mathfrak A }_{5}(0) =0\,.
\end{eqnarray*}  
Graphics for ${\cal A}_{\nu}(L)$ as functions of $\nu$ at fixed values of $L$
can be found in our last papers~\cite{BMS05}.
We compare the basic ingredients of (M)FAPT in 
Table \ref{tab:PT.APT.FAPT} with their counterparts 
in conventional PT and APT.
\begin{table}[htb]
\caption{Comparison of PT, APT, FAPT ($\Ds L=\ln\left(Q^2/\Lambda^2\right)$),
 and MFAPT ($\Ds L=\ln\left(s/\Lambda^2\right)$).
 In the row, named `Inverse powers', 
 we put ${\mathfrak A}_{-m}(L)=L^m+O(\pi^2)$
 that symbolically encodes just the item (2) of ${\mathfrak A}_{\nu}(L)$ properties,
 see the list on the previous page. 
 \label{tab:PT.APT.FAPT}\vspace*{1mm}}
  \begin{tabular}{ccccc}\hline\hline
  ~~~~~~Theory~~~~~~
          &~~~~~~\text{PT}~~~~~~
                      &~~~~~~APT~~~~~~~
                               &~~~~~~FAPT~~~~~~
                                        &~~~~~~MFAPT~~~~~~$\vphantom{^{\int}}$
  \\ \hline
   Space  & $\Big\{a^\nu\Big\}_{\nu\in\mathbb{R}}\vphantom{^{\big|}_{\big|}}$
                      & $\Big\{{\cal A}_m\Big\}_{m\in\mathbb{N}}$
                               & $\Big\{{\cal A}_\nu\Big\}_{\nu\in\mathbb{R}}$
                                        & $\Big\{{\mathfrak A}_\nu\Big\}_{\nu\in\mathbb{R}}$
  \\ \hline
  Series expansion
          & $\sum\limits_{m}f_m\,a^m(L)\vphantom{^{\big|}_{\big|}}$
                      & $\sum\limits_{m}f_m\,{\cal A}_m(L)$
                               & $\sum\limits_{m}f_m\,{\cal A}_m(L)$
                                        & $\sum\limits_{m}f_m\,{\mathfrak A}_m(L)$
  \\ \hline
  Inverse powers
          & $\left(a(L)\right)^{-m}\vphantom{^{\big|}_{\big|}}$
                      & ~\text{---}~
                               & ${\cal A}_{-m}(L)=L^m$
                                        & ${\mathfrak A}_{-m}(L)=L^m+O(\pi^2)$
  \\ \hline
  Index derivative
          & $a^{\nu}\ln^{{k}}{a}\vphantom{^{\big|}_{\big|}}   $
                      & ~\text{---}~
                               & ${\cal D}^{k}{\cal A}_\nu$
                                        & ${\cal D}^{k}\,{\mathfrak A}_{\nu}$
  \\ \hline\hline
  \end{tabular}
\end{table}

\section{Development of (M)FAPT: Two-loop coupling}
The two-loop equation for the normalized coupling $a=b_0\,\alpha/(4\pi)$ 
is 
\begin{eqnarray}
 \frac{d a_{(2)}(L)}{dL}
    = - a_{(2)}^2(L)\left[1 + c_1\,a_{(2)}(L)\right]
      \quad \text{with}~c_1\equiv\frac{b_1}{b_0^2}\,.
\end{eqnarray}
\begin{figure}[t]
 \begin{minipage}{\textwidth}
  \centerline{\includegraphics[width=0.45\textwidth]{
   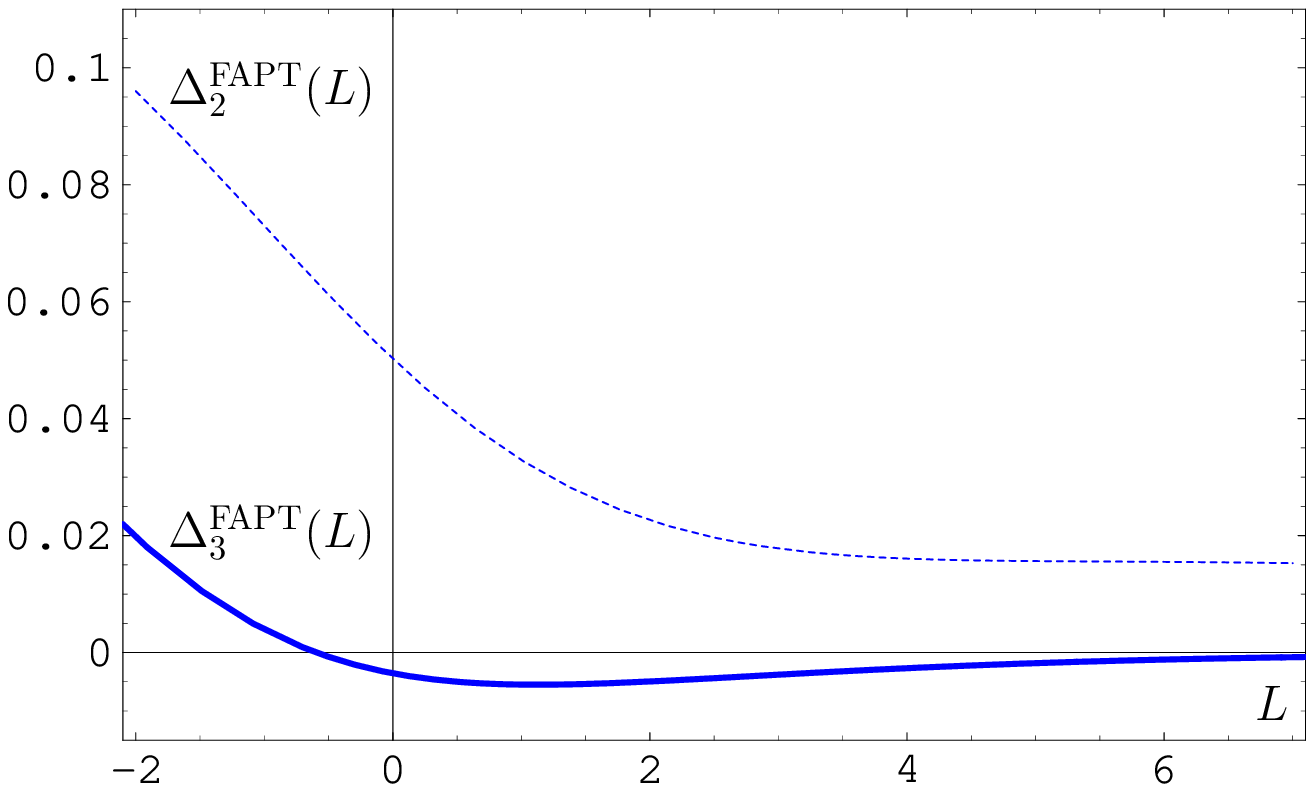}~~~%
              \includegraphics[width=0.45\textwidth]{
   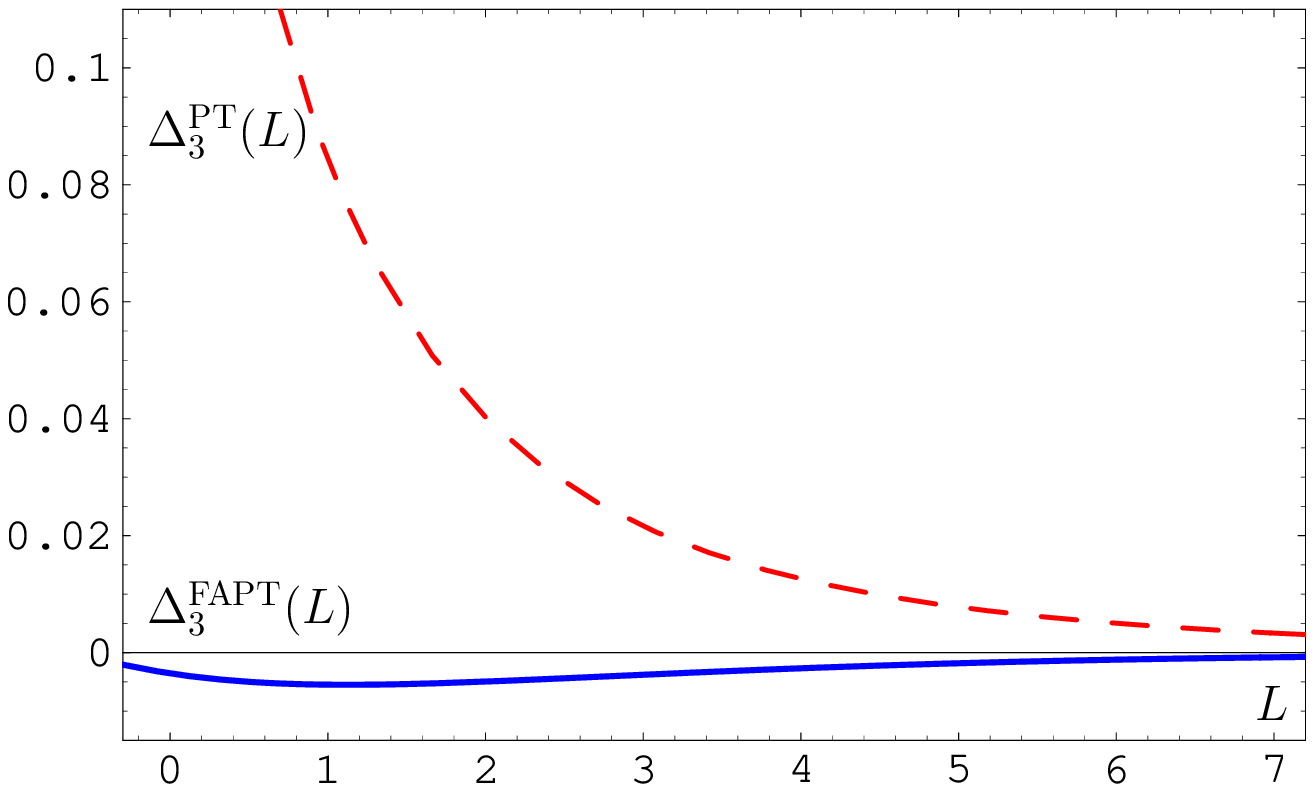}}%
  \caption{Left panel: Comparison of relative errors 
   $\Delta_2^\text{FAPT}(L)$ (dotted line) and 
   $\Delta_3^\text{FAPT}(L)$ (solid line) in FAPT.
   Right panel: Comparison of relative errors 
   $\Delta_3^\text{PT}(L)$ (dashed line) in standard PT 
   and $\Delta_3^\text{FAPT}(L)$ (solid line) in FAPT.
  \label{fig:FAPT23_PT3}}
\end{minipage}
\end{figure}
RG solution of this equation assumes the following form:
\begin{eqnarray}
 \frac{1}{a_{(2)}(L)} + c_1 \ln\left[\frac{a_{(2)}(L)}{1+c_1 a_{(2)}(L)}\right] 
  = L = \frac{1}{a_{(1)}(L)}\,.
\end{eqnarray}
We can expand $a_{(2)}(L)$ 
in terms of $a_{(1)}(L)=1/L$ 
with inclusion of terms ${\cal O}(a_{(1)}^{3})$:
\begin{eqnarray*}
 a_{(2)}(L)
   = a_{(1)}(L)
   + c_1\,a_{(1)}^2(L)\,\ln\,a_{(1)}(L)
   + c_1^2\,a_{(1)}^3(L)
        \left(\ln^2 a_{(1)}(L)
            + \ln\,a_{(1)}(L)
            -1 \right) 
       + \ldots\,.
\end{eqnarray*}
Analytic version of this expansion is
\begin{eqnarray*}\!\!\!
 {\cal A }_{1}^{(2);\text{FAPT}}(L)
  &=& {\cal A }_{1}^{(1)}
      + c_1\,{\cal D}\,{\cal A }_{\nu=2}^{(1)}
      + c_1^2\left({\cal D}^{2}+{\cal D}-1\right)
        {\cal A }_{\nu=3}^{(1)}
      + \ldots\,.
\end{eqnarray*} 
In Fig.\ \ref{fig:FAPT23_PT3} 
we demonstrate nice convergence of this expansion 
using relative errors 
of the 2- and 3-term approximations:
\begin{eqnarray*}
  \Delta_2^\text{FAPT}(L) 
    &=& 1 
     - \frac{{\cal A}_{1}^{(1)}(L) + c_1\,{\cal D}{\cal A}_{\nu=2}^{(1)}(L)}
            {{\cal A}_1^{(2)}(L)}\,;\\
  \Delta_3^\text{FAPT}(L) 
    &=& \Delta_2^\text{FAPT}(L)
     - \frac{c_1^2\,\left({\cal D}^{2}+{\cal D}-1\right)\,{\cal A }_{\nu=3}^{(1)}(L)}
            {{\cal A}_1^{(2)}(L)}\,;
\end{eqnarray*}
\begin{eqnarray*}
  \Delta_3^\text{PT}(L) 
    = 1
     - \frac{a_{(1)}(L)
           + c_1\,a_{(1)}^2(L)\,\ln\,a_{(1)}(L)
           + c_1^2\,a_{(1)}^3(L)
             \left(\ln^2 a_{(1)}(L) + \ln\,a_{(1)}(L) - 1\right)}
            {a_{(2)}(L)}\,.
\end{eqnarray*}
We see that relative accuracy of the 3-term approximation in FAPT
(see the left panel of Fig.\ \ref{fig:FAPT23_PT3}) is better 
than 2\% for $L\geq-2$.
In the same time, the right panel of Fig.\ \ref{fig:FAPT23_PT3}
demonstrates 
that relative accuracy of the same 3-term approximation in standard PT
even at $L\approx1$ is much higher --- about 10\%,
whereas in FAPT it is smaller than 1\%!

\begin{figure}[hb]
 \begin{minipage}{\textwidth}
  \centerline{\includegraphics[width=0.45\textwidth]{
   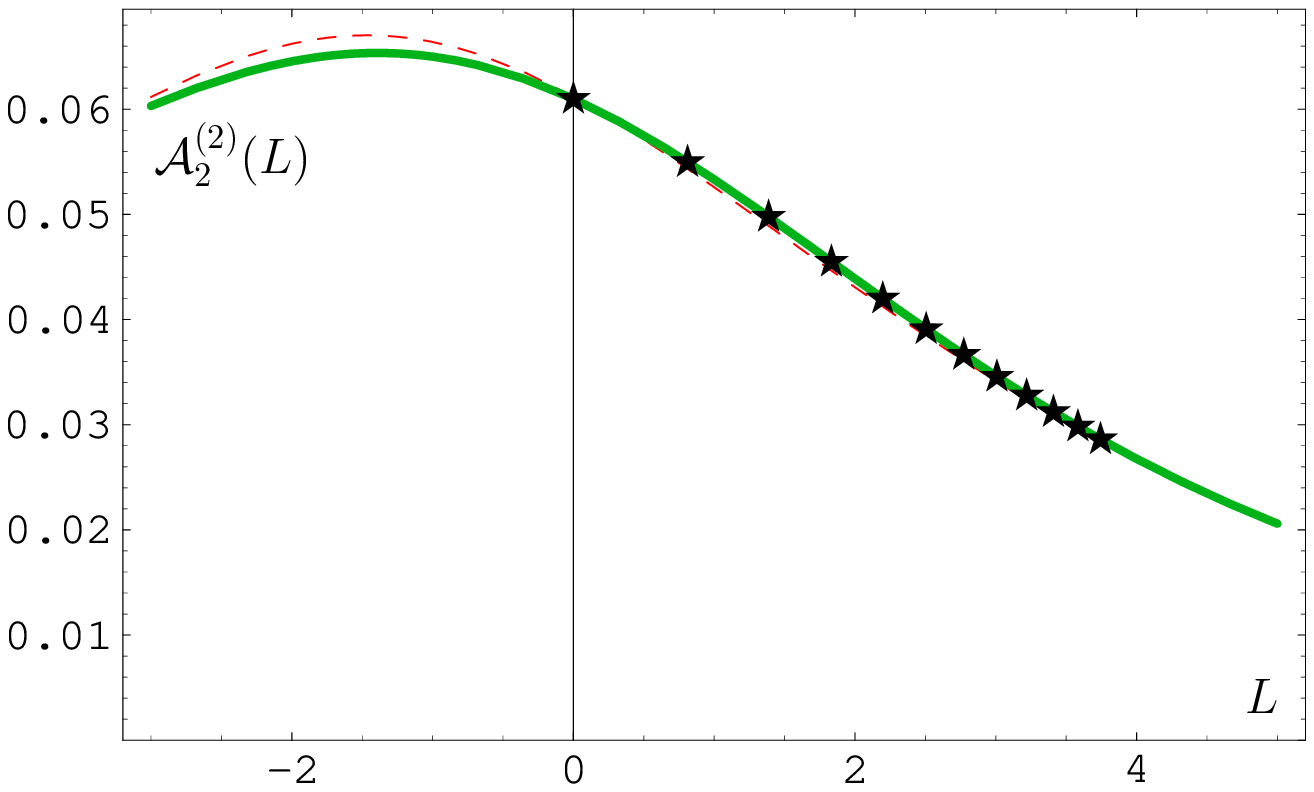}~~~%
              \includegraphics[width=0.45\textwidth]{
   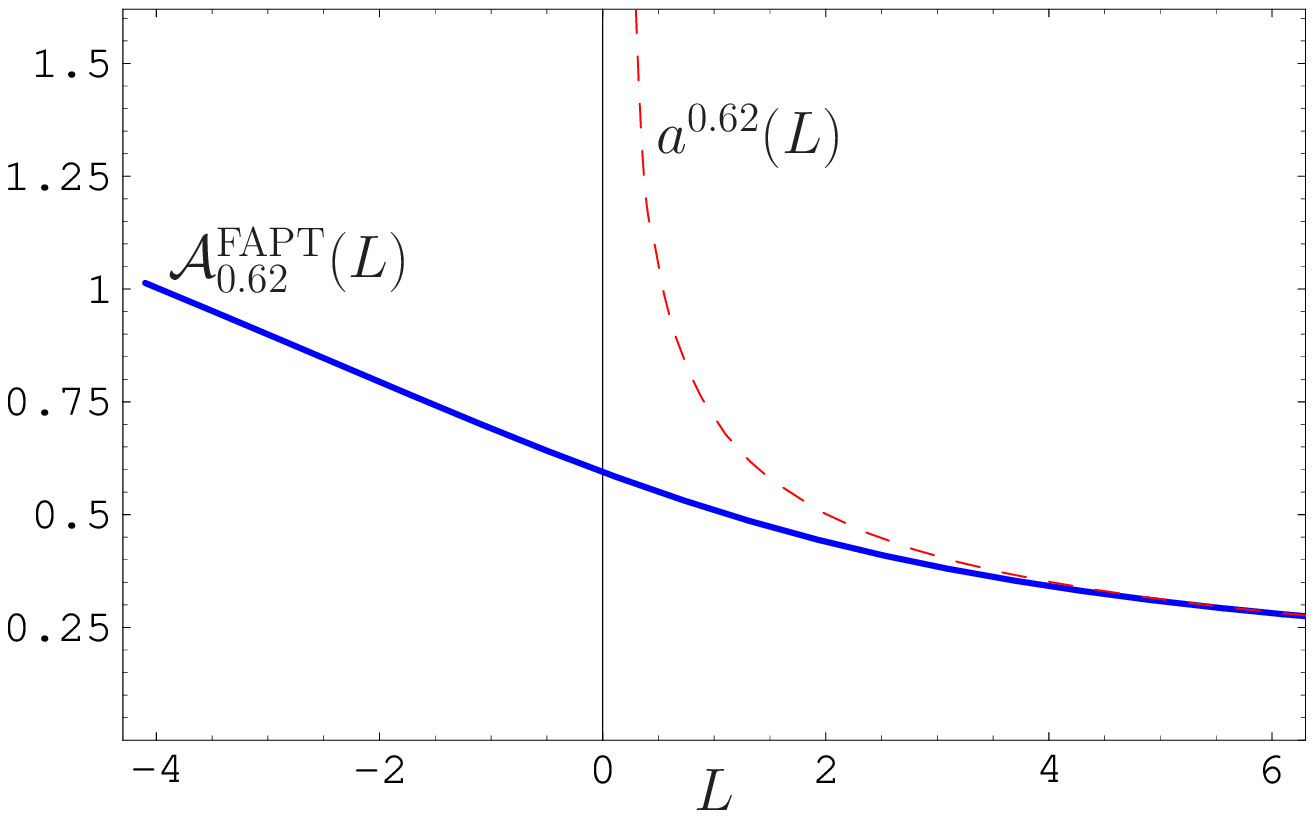}\vspace*{-1mm}}%
  \caption{{Left panel}: The solid line corresponds to ${\cal A}_2^{(2)}(L)$,
  computed analytically via Eq.\ (\ref{eq:A2_FAPT3});
  dashed line represents the result of a numerical integration,
  while stars correspond to the available numerical results 
  of Magradze in \cite{Mag03u}.
  {Right panel}: The solid line represents ${\cal A}_{0.62}^{(2);\text{FAPT}}(L)$,
  computed analytically via Eq.\ (\ref{eq:A2_FAPT3}), while
  the dashed line stands for $a^{0.62}_{(2)}(L)$.
  \label{fig:Fapt-Pt.A2}\vspace*{-3mm}}
\end{minipage}
\end{figure}

We can also obtain the corresponding expansion for the two-loop coupling with
index $\nu$:
\begin{eqnarray}\!\!\!
 {\cal A}_{\nu}^{(2);\text{FAPT}}(L)
   &=& {\cal A }_{\nu}^{(1)}(L)
         + c_1\nu\,{\cal D}\,{\cal A }_{\nu+1}^{(1)}(L)
         + c_1^2\nu\left[\frac{\nu+1}{2}{\cal D}^{2}\!+\!{\cal D}\!-\!1\right]
              \!{\cal A }_{\nu+2}^{(1)}(L)
         + \ldots\,.~~~\label{eq:A2_FAPT3}
\end{eqnarray}
and display comparison of different results for ${\cal A}_2^{(2);\text{FAPT}}(L)$
on the left panel of Fig.\ \ref{fig:Fapt-Pt.A2}.
On the right panel of this figure 
we show  comparison of FAPT and standard QCD PT with respect
to the fractional index (power) of the coupling,
fixed at the value $\nu=0.62$.

In Minkowskian region convergence of MFAPT expansion
for the two-loop coupling 
\begin{eqnarray}\!\!\!
 {\mathfrak A}_{2}^{(2);\text{MFAPT}}(L)
  &=& {\mathfrak A }_{2}^{(1)}(L)
       + 2\,c_1\,{\cal D}\,{\mathfrak A }_{\nu=3}^{(1)}(L)
       + c_1^2\,\left[3\,{\cal D}^{2}+2\,{\cal D}-2\right]
         \!{\mathfrak A }_{\nu=4}^{(1)}(L)
       + \ldots\,.~~~
 \label{eq:MFAPT.U23}
\end{eqnarray} 
is also nice,
but in the vicinity of the point $L=0$ 
(Landau pole in the standard PT)
it is not so fast,
so that we need to take into account
$O(c_1^5)$-terms in order to reach 5\% level of accuracy,
for more details look in~\cite{BMS05}.

\section{Electromagnetic pion form factor at NLO}
 Scaled hard-scattering amplitude truncated at the next-to-leading order 
(NLO) 
and evaluated at renormalization scale $\mu_{R}^2=\lambda_{R} Q^2$ 
reads~
\cite{FGOC81,DR81,BT87,MNP99a}
\begin{eqnarray}
  T^\text{NLO}_{H}\left(x ,y;\mu_{F}^2,Q^2\right)
  = \frac{\alpha_s\left(\lambda_{R}Q^2\right)}{Q^2}\,
     t_{H}^{(0)}(x,y)
  + \frac{\alpha_s^2\left(\lambda_{R}Q^2\right)}{4\pi\,Q^2}\,
      t_{H}^{(1)}(x,y;{\mu_{F}^2}/{Q^2})~~~\label{eq:T_Hard_NLO}  
\end{eqnarray}
with shorthand notation ($\bar{x}\equiv1-x$)
\begin{eqnarray*}
 ~~t_{H}^{(1)}(x,y;{\mu_{F}^2}/{Q^2})
  = \left[C_{F}\,t_{\text{H}}^{(0)}\left(x,y\right)
     \left[2\Big(3+\ln\,(\bar{x}\bar{y})\Big)
      \ln\frac{Q^2}{\mu_{F}^2}\right]
        + b_0\,t_{H}^{(1,\beta)}\left(x,y;\lambda_{R}\right)
        + t_{H}^{(\text{FG})}(x,y)
    \right]\,.
\end{eqnarray*}
The leading twist-2 pion distribution amplitude (DA)~\cite{Rad77} 
at normalization scale $\mu_F^2$ 
is given by~\cite{ER80}
\begin{eqnarray*}
 \varphi_\pi(x,\mu_F^2)
  =  6\,x\,(1-x)
      \left[ 1
         + a_2(\mu_F^2) \, C_2^{3/2}(2 x -1)
         + a_4(\mu_F^2) \, C_4^{3/2}(2 x -1) 
         + \ldots\,
      \right]\,.
\end{eqnarray*}
All {nonperturbative} information is encapsulated in Gegenbauer coefficients 
$a_n(\mu^2_F)$.

To obtain factorized part of pion form factor (FF) one needs to convolute
the pion DA with the hard-scattering amplitude:
\begin{eqnarray*}
 F_\pi^\text{Fact}(Q^2) 
  = \varphi_\pi(x;\mu_{F}^2)\convo{x}
      T^\text{NLO}_{H}\left(x,y;\mu_{F}^2,Q^2\right)
         \convo{y}\varphi_\pi(y;\mu_{F}^2)\,.
\end{eqnarray*}
In order to obtain the analytic expression for the pion FF at NLO 
in~\cite{SSK99,SSK00} the so-called ``Naive Analytization'' 
has been suggested.
It uses analytic image only for coupling itself, ${\cal A}_{1}^{(2)}$,
but not for its powers.
In contrast and in full accord with the APT ideology 
the receipt of ``Maximal Analytization'' has been proposed
recently in~\cite{BPSS04}. 
The corresponding expressions for the analytized hard amplitudes
read as follows:
\begin{eqnarray*}
 \left[Q^2 T_{H}\left(x,y,Q^2\right)
 \right]_\text{Nai-An} 
 &=& {\cal A}_{1}^{(2)}(\lambda_{R} Q^2)\,t_{H}^{(0)}(x,y)
 + \frac{\left({\cal A}_{1}^{(2)}(\lambda_{R} Q^2)\right)^2}{4\pi}\,
    t_{H}^{(1)}\left(x,y;\lambda_{R},\frac{\mu_{F}^2}{Q^2}\right)\,;\\
 \left[Q^2 T_{H}\left(x,y,Q^2\right)
 \right]_\text{Max-An}
 &=& {\cal A}_{1}^{(2)}(\lambda_{R} Q^2)\,t_{H}^{(0)}(x,y)
    + \frac{\acal_{2}^{(2)}(\lambda_{R} Q^2)}{4\pi}\,
       t_{H}^{(1)}\left(x,y;\lambda_{R},\frac{\mu_{F}^2}{Q^2}\right)\,.
\end{eqnarray*}
\begin{figure}[h]
 \begin{minipage}{\textwidth}
  \centerline{\includegraphics[width=0.325\textwidth]{
   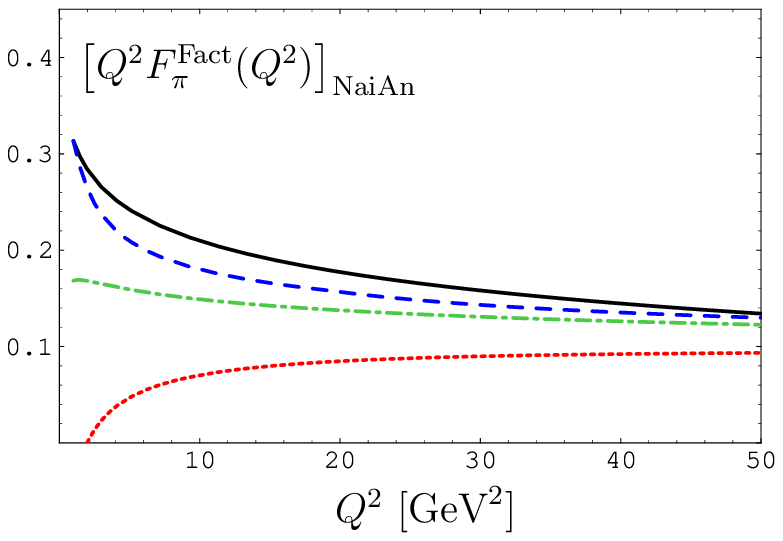}~%
              \includegraphics[width=0.325\textwidth]{
   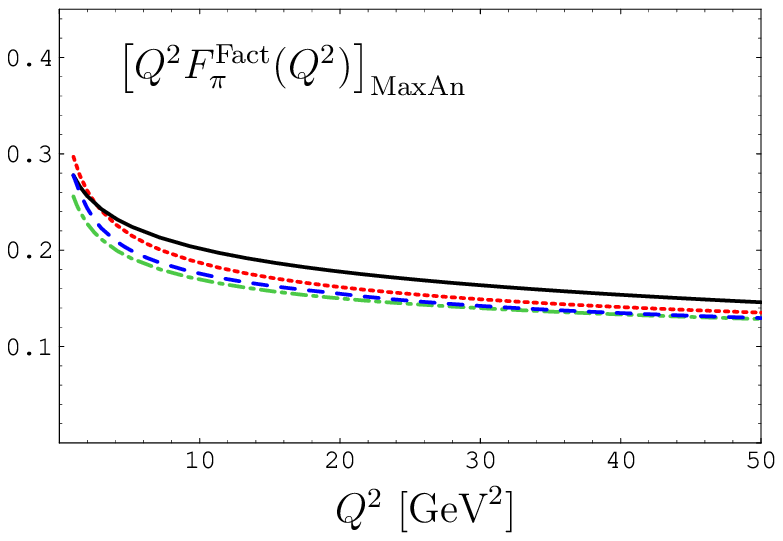}~%
              \includegraphics[width=0.325\textwidth]{
   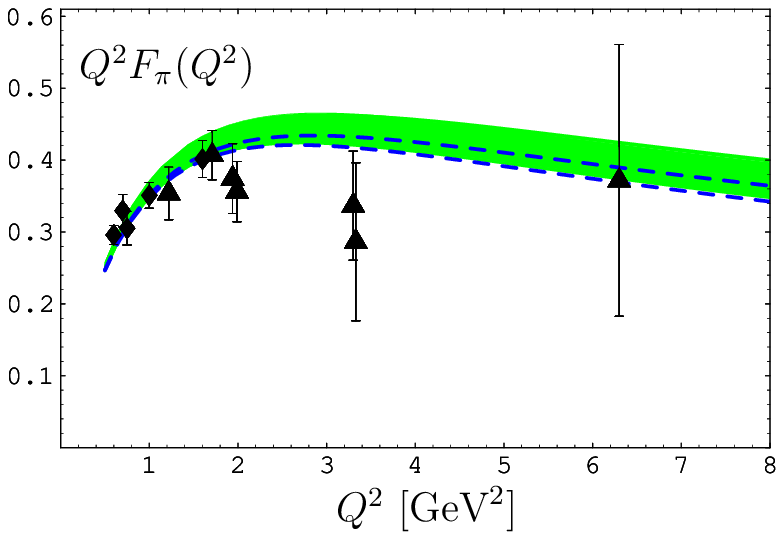}}
  \caption{{Left panel}: Factorized pion FF in the ``Naive Analytization''.
   {Central panel}: Factorized pion FF in the ``Maximal Analytization''.
   On both panels solid lines correspond to the scale setting $\mu_R^2=1$~GeV$^2$, 
   dashed lines --- to $\mu_R^2=Q^2$, dotted lines --- to the BLM prescription,
   whereas dash-dotted lines --- to the $\alpha_v$-scheme.
   {Right panel}: Predictions for the scaled pion form factor calculated with 
  the BMS\ bunch of the pion DAs.
  The dashed lines inside the strip indicate the corresponding area 
  of predictions obtained with the asymptotic pion DA.
  The experimental data are taken from \protect{\cite{JLAB00}}
  (diamonds) and \cite{FFPI73}, \cite{FFPI76} (triangles).
     \label{fig:PionFF_Nai_Max_Strip}\vspace*{-3mm}}
\end{minipage}
\end{figure}
In Fig.\ \ref{fig:PionFF_Nai_Max_Strip} we show the predictions for
the factorized pion FF in the ``Naive'' and in the ``Maximal Analytization''
approaches. 
We see that in the ``Maximal Analytization'' approach
the obtained results are practically insensitive 
to the renormalization scheme and scale-setting choice
(already at the NLO level).

We show also the graphics for the whole pion FF,
obtained in APT with the ``Maximally Analytic'' procedure
using the Ward identity to match the non-factorized and factorized parts
of the pion FF, see the right panel of Fig.\ \ref{fig:PionFF_Nai_Max_Strip}.
The green strip in this figure contains both nonperturbative uncertainties 
from nonlocal QCD sum rules~\cite{BMS01,BP06,AB06parus} 
and renormalization scheme and scale ambiguities 
at the level of the NLO accuracy.

It is interesting to note here that FAPT approach,
used in~\cite{BKS05} for analytization of the $\ln(Q^2/\mu_{F}^2)$-terms 
in the hard amplitude (\ref{eq:T_Hard_NLO}),
diminishes also the dependence on the factorization scale setting 
in the interval $\mu_{F}^2=1-10$~GeV$^2$.

\section{Concluding Remarks}
We conclude with the following resume:\\
\ding{172} The implementation of the analyticity concept
    (the dispersion relations) from the level of the coupling 
     and its powers to the level of QCD amplitudes as a whole
     generates extension of the APT to (M)FAPT\,;\\
\ding{173} We formulate the rules how to apply (M)FAPT at the two- and three-loop levels;\\
\ding{174} We show that convergence of the perturbative expansion 
      is significantly improved when using non-power (  M)FAPT expansion;\\
\ding{175} As an additional advantage we obtain the minimal sensitivity 
     to both the renormalization and factorization scale setting,
     revealed on the example of the pion electromagnetic form factor.

\bigskip

\textbf{Acknowledgments:}
This investigation was supported in part 
by the Deutsche Forschungsgemeinschaft
(Projects DFG 436 RUS 113/881/0),
the Heisenberg--Landau Programme, grant 2007, 
the Russian Foundation for Fundamental Research, 
grants No.\ 05-01-00992 and 07-02-91557,
and the BRFBR--JINR Cooperation Programme, contract No.\ F06D-002.
   



\end{document}